\documentclass[aps,prb,twocolumn,superscriptaddress,showpacs]{revtex4}

\usepackage{amsmath}
\usepackage{amssymb}
\usepackage{epsfig}

\begin{document}

\title{Structure, phase behavior and inhomogeneous fluid properties of
binary dendrimer mixtures}

\author{I.~O.~G\"otze}
\affiliation{Institut f{\"u}r Theoretische Physik II,
Heinrich-Heine-Universit{\"a}t D{\"u}sseldorf, Universit{\"a}tsstra{\ss}e
1, D-40225 D{\"u}sseldorf, Germany}

\author{A.~J.~Archer}
\affiliation{H.~H.~Wills Physics Laboratory, University of Bristol,
Bristol BS8 1TL, UK}

\author{C.~N.~Likos}
\affiliation{Institut f{\"u}r Theoretische Physik II,
Heinrich-Heine-Universit{\"a}t D{\"u}sseldorf, Universit{\"a}tsstra{\ss}e
1, D-40225 D{\"u}sseldorf, Germany}

\date{\today}

\begin{abstract}
The effective pair potentials between different kinds of dendrimers in
solution can be well approximated by appropriate Gaussian functions.
We find that in binary dendrimer mixtures the range and strength 
of the effective interactions
depend strongly upon the specific dendrimer architecture. 
We consider two different types of dendrimer mixtures, employing the Gaussian 
effective pair potentials,
to determine the bulk fluid structure and phase behavior. Using a simple
mean field density functional theory (DFT) we find good agreement between
theory and simulation results for the bulk fluid structure. Depending on the mixture, we find bulk fluid-fluid phase separation
(macro-phase separation) or micro-phase
separation, i.e., a transition to a state characterized by undamped
periodic concentration fluctuations. We also determine the inhomogeneous
fluid structure for confinement in spherical cavities. Again, we find good
agreement between the DFT and simulation results. For the
dendrimer mixture exhibiting micro-phase separation, we observe rather
striking pattern formation under confinement.
\end{abstract}

\pacs{61.25.Hq, 82.70.Dd, 36.20.Ey, 61.12.Ex}

\maketitle

\section{Introduction}

Dendrimers are highly branched polymeric macromolecules. They are
synthesized from monomers that can form bonds with at least three other
monomers. Starting from a suitable central molecule, a number of
generations of monomers are sequentially added. With the addition of
each successive generation, the dendrimer becomes increasingly branched in
its internal structure.\cite{newcome:96,fischer:99,bosman:99,
tomalia:01,likos:04}
A dendrimer with $n$ generations
of monomers is denoted a G$n$ dendrimer. Dendrimers have numerous
technological applications -- for example in the targeted delivery of 
drugs\cite{lee,boas:04} and, in particular, anti-cancer 
agents,\cite{quintana:02,kohle:03,choi:05} 
in light-harvesting applications,\cite{light-harvesting}
and also as a tool for the 
development of nonviral gene delivery.\cite{goessl:02} 
In a dendrimer solution, the
individual macromolecules feature different conformations due to
fluctuations of the monomers. However, statistically one finds that the
intra-molecular monomer distribution is almost spherically symmetric.
For the statistical behavior of the ensemble of macromolecules only the
average over all conformations
and orientations of the interacting dendrimers has to be taken into
account, provided no phases with orientational order appear. This suggests
treating the interaction between dendrimers via radially
symmetric effective pair potentials, even though a dendrimer is a complex,
structured object. Also, given the number of degrees of freedom in
each dendrimer, determining the structure and phase behavior of
suspensions of dendrimers starting from an atomistic view point is a
tough problem. Obtaining an effective interaction potential between
pairs of dendrimers (e.g.~between the centres of mass) is a way to
surmount this. In such a
procedure, one is effectively integrating in the partition function over
the internal monomeric degrees of freedom of each dendrimer, resulting
in an effective Hamiltonian for the dendrimer mixture that treats each
dendrimer as a point particle.\cite{habil, ingo:jcp:04,
likos:rosenfeldt:02, ingo:macrom, GotzeLikosJPCM2005} However, in
principle such an effective Hamiltonian also includes effective
three-body and higher body effective interactions between the
particles.\cite{habil,GotzeLikosJPCM2005} For dendrimers in solution one
finds that these three and higher body interactions are small compared to
the two-body potentials,\cite{GotzeLikosJPCM2005} so neglecting these
higher body terms in the effective Hamiltonian is justified.
Given an effective Hamiltonian for point particles one can then determine
the fluid structure and phase behavior by means of theory or
simulations. Thus,
effective interactions are a powerful tool to bridge the length scales
and to extract macroscopic properties of dendrimer solutions
from microscopic details.

In this paper we use such an approach to study binary dendrimer mixtures.
We find that all effective pair potentials between dendrimers take a
Gaussian form.\cite{ingo:jcp:04,lang:jpcm:00} 
The parameters in these potentials depend strongly
on the specific dendrimer architectures, to the extent that we can `tune'
their range and strength by changing the number of generations
and flexibility between monomers in the dendrimers.
This enables us to study the occurrence of a number of phenomena that
have been predicted for Gaussian mixtures, such as bulk fluid-fluid
demixing\cite{ard:pre:00,andy1} and micro-phase separation.\cite{andy2}
Micro-phase separation\cite{sear:99} is characterized by a transition
in the fluid to a state with undamped periodic concentration fluctuations
which is thought to indicate a thermodynamic transition to a rather
unusual crystalline state.\cite{andy2} When Gaussian mixtures are 
subject to geometrical confinement, 
pattern formation can spontaneously arise.\cite{andy2} We find dendrimer
mixtures exhibiting all these phenomena.
Thus, we are able to demonstrate that the rich phenomenology encountered 
in Gaussian mixtures corresponds to {\it real}
systems.

This paper is laid out as follows: In Sec.~\ref{sec:eff_pair_pots} we
describe our procedure for determining the effective pair potentials in
dendrimer mixtures and give results for two particular systems. In
Sec.~\ref{sec:theory} we briefly describe theories that apply for Gaussian
mixtures and then in Sec.~\ref{sec:results} we present theory and
simulation results for the bulk structure and phase behavior of
particular dendrimer mixtures. In Sec.~\ref{sec:confined} we present
theory and simulation results for two particular dendrimer mixtures
confined in spherical cavities. Finally, in Sec.~\ref{sec:conclusions} we
draw our conclusions.

\section{Effective pair potentials}
\label{sec:eff_pair_pots}

It has been shown by means of monomer-resolved simulations and theory,
that the effective pair potentials between the centres of mass of
dendrimers in a good solvent
can be modelled by a purely repulsive Gaussian
potential of the form:\cite{ingo:jcp:04}
$V(r)=\epsilon \exp(-r^2/{\rm R}^2)$,
where the parameters $\epsilon>0$ and ${\rm R}$ depend strongly upon the
parameters characterizing the dendrimer's specific
architecture. 
In particular, they are both affected by the terminal generation number 
and the bond length between successive generations.
It is therefore possible to `tune' the effective
interactions by modifying these parameters.
This result has also been confirmed by making
comparison with experimental results.\cite{likos:rosenfeldt:02} In the
present work we determine the effective pair potential between {\em
different} types of dendrimers. We find that in binary mixtures,
these can also be well approximated by Gaussian
functions of different ranges and strengths.
The pair potential between two dendrimers of
species $i$ and $j$ thus takes the form:
\begin{equation}
V_{ij}(r)=\epsilon_{ij} \exp(-r^2/{\rm R}_{ij}^2),
\label{eq:V_ij}
\end{equation}
where the parameters $\epsilon_{ij}>0$ and ${\rm R}_{ij}$ depend strongly on
the specific properties of the two types of dendrimers in question.

In order to obtain the effective interaction potentials between a pair of
dendrimers,
we employed a simple coarse-grained model, the so-called bead-thread
model,\cite{ingo:macrom} which treats the monomers as hard spheres of
diameter $\sigma$, connected by ideal threads of maximum extension
$(1+\delta)\sigma$.
The parameter $\delta$ corresponds to the length of the polymer
chains between two branching points in the dendrimer.
Increasing the number of chemical bonds between two branching points leads
to an increase in flexibility of the polymer and thus to softer but more
long ranged effective interactions between dendrimers.
Large values of $\delta$ are used to model dendrimers with open
structures and compact dendrimers are modelled by choosing a
smaller value of $\delta$. As the number of generations (monomers) is
increased, the radius of the molecule, of course, increases and the
average monomer density inside the molecule grows as well.
For a fixed number of monomers in the dendrimer, the value of the
effective pair potential for complete overlap of the centres of mass of a
pair of dendrimers decreases upon increasing $\delta$.
Upon increasing the generation number $G$, keeping $\delta$
fixed, the effective interaction becomes more repulsive because of the
exponential growth of the monomer number with $G$.
By varying $\delta$ and $G$, one can easily modify the dendrimer
architecture and thereby systematically `tune' the effective pair
potential between dendrimers of the same kind.\cite{ingo:jcp:04}
However, it is much less
straightforward to a priori determine the cross-interaction between
two dendrimers with different values for the parameters $\delta$ and $G$.
At the same time, employing mixtures of different dendrimer types dramatically
increases the freedom to tailor the macroscopic properties of the system.

\begin{figure}
  \begin{center}
  \includegraphics[width=8cm,clip]{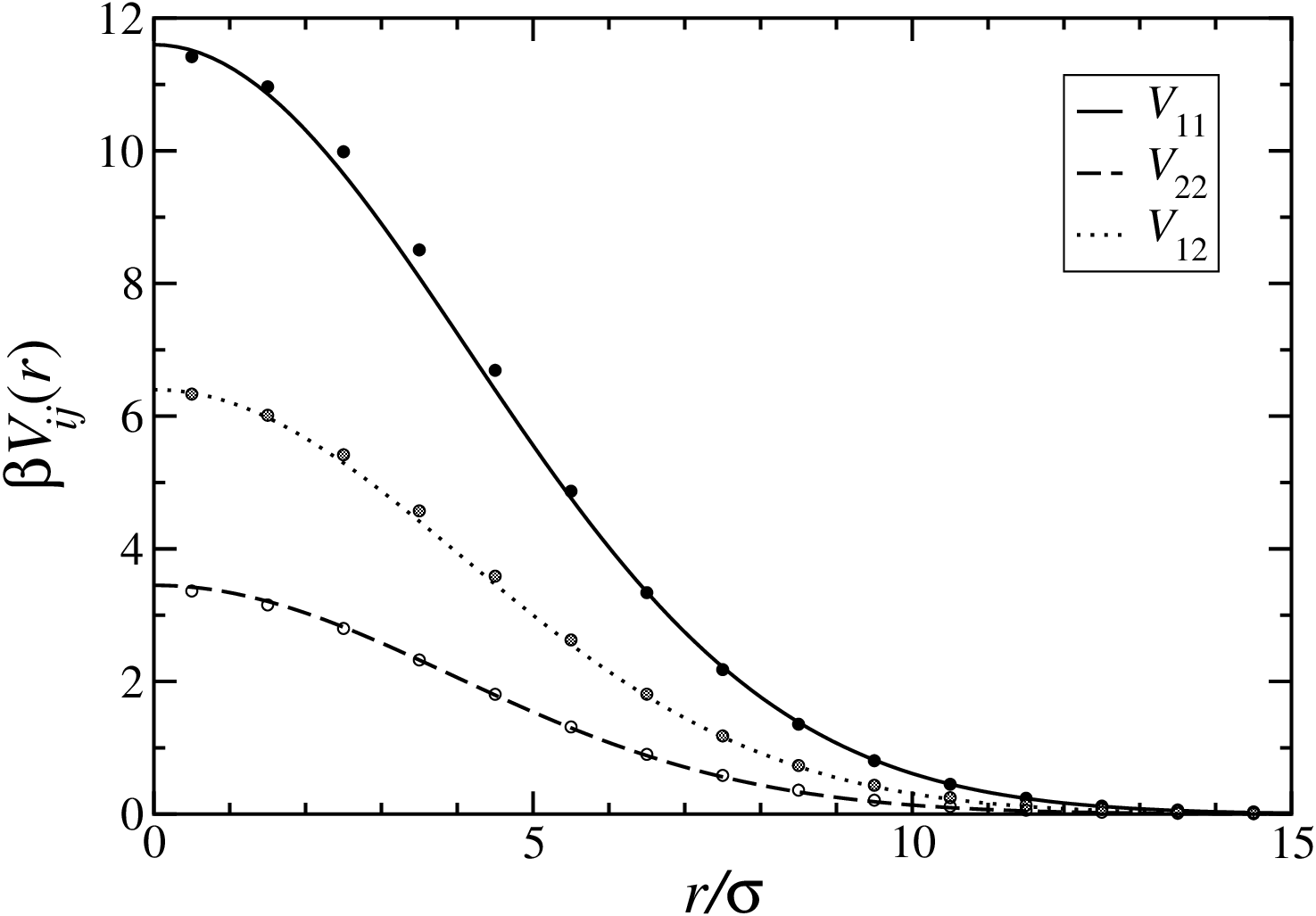}
  \end{center}
  \caption{Effective pair potentials $V_{ij}(r)$ between
  dendrimers in a binary mixture exhibiting macro-phase separation
  (system A).
  Species 1 is a G4 dendrimers with $\delta=2.0\sigma$ and species 2 is
  a G3 dendrimers with $\delta=3.0\sigma$. 
  The corresponding MC data are shown as circles.}
  \label{macro:pot}
\end{figure}

\begin{figure}
  \begin{center}
  \includegraphics[width=8cm,clip]{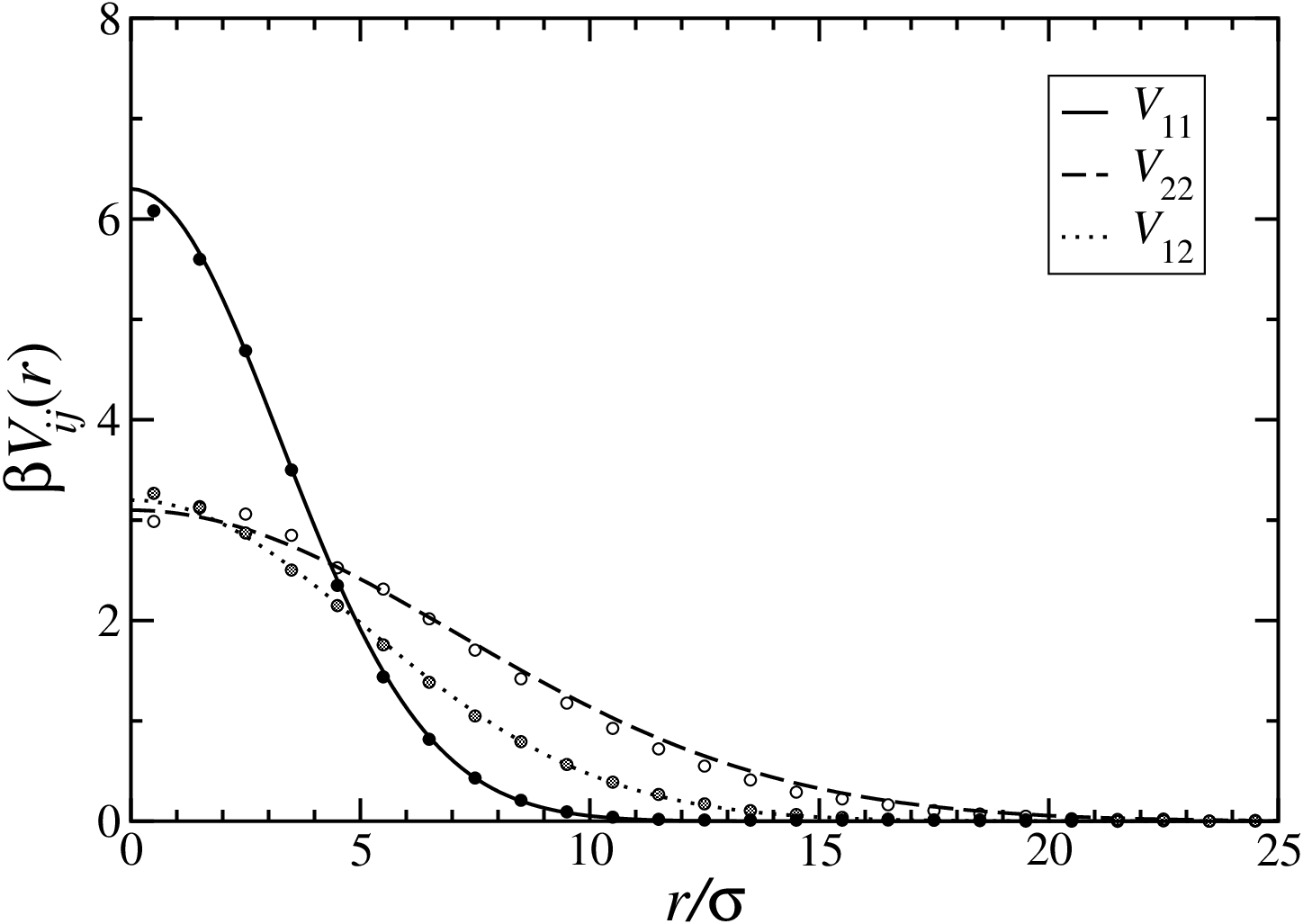}
  \end{center}
  \caption{Effective pair potentials $V_{ij}(r)$ between
  dendrimers in a binary mixture exhibiting micro-phase separation
  (system B).
  Species 1 is a G3 dendrimers with $\delta=2.0\sigma$ and species 2 is
  a G4 dendrimers with $\delta=4.7\sigma$. The corresponding MC data are shown as circles.}
  \label{micro:pot}
\end{figure}

To determine effective dendrimer interactions $V_{ij}(r), i,j=1,2$ in a 
binary mixture, we employ monomer-resolved Monte-Carlo (MC)
simulations of the bead-thread model for
dendrimers of different generations and varying thread length $\delta$.
We simulate pairs
of dendrimers of species $i$ and $j$ and determine in the simulation
the probability $P_{ij}(r)$ of finding their centers of mass separated by
distance $r$. Thereafter, the effective potential $V_{ij}(r)$ 
can be obtained using the relation:
\begin{equation}
\beta V_{ij}(r) = -\ln[g_{ij}(r)],
\end{equation}
where $\beta=1/k_BT$ is the inverse temperature and 
$g_{ij}(r) \propto P_{ij}(r)$, are the
partial radial distribution functions.\cite{habil} 
Due to the strong repulsion for complete overlap between dendrimers,
we use biased MC, as described in Ref.~\onlinecite{ingo:jcp:04},
to obtain the effective potentials.
The MC-results for $V_{ij}(r)$ can be very well described by Gaussian
functions of the form in Eq.~(\ref{eq:V_ij}), as can be 
seen in Figs.\ \ref{macro:pot} and \ref{micro:pot}.

Having derived effective Gaussian pair potentials for a given binary
dendrimer mixture it is then possible to apply theories developed for
point particles to calculate the mixture structure and thermodynamics.
Such a theory is described in the next section. Typically, we find that
most binary dendrimer mixtures exhibit bulk fluid-fluid phase separation 
(macro-phase separation). In Fig.~\ref{macro:pot} we display the effective pair
potentials for a particular binary mixture that does exhibit macro-phase
separation. This mixture is composed of G4 dendrimers with
$\delta=2.0\sigma$ together with G3 dendrimers with $\delta=3.0\sigma$.
However, in one particular mixture that we investigated, we find that the
mixture exhibits micro-phase separation.\cite{andy2} This mixture is
composed of G3 dendrimers with $\delta=2.0\sigma$ together with G4
dendrimers with $\delta=4.7\sigma$. The effective pair potentials for this
mixture are displayed in Fig.~\ref{micro:pot}. It can be seen there that 
the key ingredient for generating micro-phase separation is
that the effective potentials in one species are short-ranged and
strongly repulsive, whilst the effective interactions in the other species
are long-ranged with soft repulsion.

\section{Theory}
\label{sec:theory}

Having derived effective pair potentials between the different species of
dendrimers in a particular binary mixture, we now use these potentials as
input to liquid state theories developed for determining the fluid
structure and thermodynamics. It has been shown that a simple mean field
DFT which generates the random phase approximation (RPA) for the fluid
direct pair correlation functions\cite{evans} provides an
accurate description of the bulk and inhomogeneous structure of fluids
composed of Gaussian particles,\cite{lang:jpcm:00,ard:pre:00,andy1}
and more generally,
for fluids composed of particles interacting via bounded or even
weakly diverging soft potentials.\cite{likos:lang,andy1b,acedo}
The intrinsic Helmholtz free energy functional of the
inhomogeneous mixture is given by:
\begin{equation}
\mathcal{F}[\{\rho_i\}]=\mathcal{F}_{\rm id}[\{\rho_i\}]
+\mathcal{F}_{\rm ex}[\{\rho_i\}],
\end{equation}
where
\begin{equation}
\mathcal{F}_{\rm id}[\{\rho_i({\bf r})\}] =k_BT \sum_i \int {\rm d}{\bf r}
\rho_i({\bf r})[\ln(\Lambda_i^3\rho_i({\bf r}))-1],
\label{eq:F_id}
\end{equation}
is the (exact) ideal gas contribution to the free energy\cite{evans} 
and
$\Lambda_i$ is the thermal de
Broglie wavelength 
for the dendrimers of species $i$. The quantity
$\mathcal{F}_{\rm ex}[\{\rho_i\}]$ is the excess (over ideal) contribution
to the free energy due to interactions between the particles. The RPA
approximation for this quantity is
\begin{equation}
\mathcal{F}_{\rm ex}[\{\rho_i\}]=\frac{1}{2}\sum_{ij}\int {\rm d}{\bf r}_1
\int {\rm d}{\bf r}_2 \, \rho_i({\bf r}_1) \rho_j({\bf r}_2) V_{ij}(|{\bf
r}_1-{\bf r}_2|).
\label{DFTRPA}
\end{equation}
The fluid direct pair correlation functions generated by the RPA
functional are simply,\cite{evans}
\begin{equation}
c_{ij}^{(2)}({\bf r}_1,{\bf r}_2)\equiv
- \frac{\beta \delta^2 \mathcal{F}_{\rm
ex}}{\delta\rho_i({\bf r}_1) \delta\rho_j({\bf r}_2)}
= -\beta V_{ij}(|{\bf r}_1-{\bf r}_2|).
\label{RPAclosure}
\end{equation}
For a given set of one body external potentials $\{V^{\rm ext}_i({\bf
r})\}$, the fluid one body density profiles are obtained by minimising the
grand potential functional:\cite{evans,andy2}
\begin{equation}
\Omega[\{\rho_i\}]=\mathcal{F}[\{\rho_i\}]-\sum_{i=1}^{2} \int {\rm d}{\bf
r}(\mu_i-V^{\rm ext}_i({\bf r}))\rho_i({\bf r}),
\label{DFTgrandpot}
\end{equation}
where $\mu_i$ is the chemical potential for species $i$.

Given such a DFT, one can calculate the fluid partial radial distribution
functions $g_{ij}(r)$ using the so called `test-particle route', in which
one fixes a particle of species $j$ and then
calculates the fluid one-body density profiles around this fixed
particle -- i.e.~one sets the external potentials
in Eq.~(\ref{DFTgrandpot}) equal to the fluid pair potentials.
The partial radial distribution function are then given by
$g_{ij}(r)=\rho_i(r)/\rho_i^{\rm b}$, where $\rho_i^{\rm b}$ is the
bulk fluid density of species $i$.

An alternative route to determining the bulk fluid structure is via the
Ornstein-Zernike (OZ) equations.\cite{hansen:book:86} In Fourier space,
the OZ equations for a two-component fluid take the form
\begin{equation}
\hat{h}_{ij}(q)=\frac{N_{ij}(q)}{D(q)},
\label{eq:OZ}
\end{equation}
where $\hat{h}_{ij}(q)$ is the (3 dimensional) Fourier transform (FT) of
$h_{ij}(r)=g_{ij}(r)-1$ and the numerator functions are
\begin{eqnarray}
N_{11}(q) &=&
\hat{c}_{11}(q)+\rho_2^{\rm
b}[\hat{c}^2_{12}(q)-\hat{c}_{11}(q)\hat{c}_{22}(q)], \nonumber\\
N_{22}(q) &=&
\hat{c}_{22}(q)+\rho_1^{\rm
b}[\hat{c}^2_{12}(q)-\hat{c}_{11}(q)\hat{c}_{22}(q)], \\
N_{12}(q) &=& \hat{c}_{12}(q), \nonumber
\end{eqnarray}
and the denominator
\begin{equation}
D(q)=[1-\rho_1^{\rm b}\hat{c}_{11}(q)][1-\rho_2^{\rm b}\hat{c}_{22}(q)]
-\rho_1^{\rm b}\rho_2^{\rm b}\hat{c}^2_{12}(q),
\label{Dq}
\end{equation}
where $\hat{c}_{ij}(q)$ is the FT of the bulk direct pair correlation
function $c_{ij}(r)$.\cite{andy1} We define the fluid partial structure
factors as follows:
\begin{equation}
S_{ij}(q)=\delta_{ij}+\sqrt{\rho_i^{\rm b}\rho_j^{\rm b}}\hat{h}_{ij}(q).
\label{eq:S_q}
\end{equation}
In Sec.~\ref{sec:results} we display some typical results for the fluid
partial radial distribution functions and structure factors. However,
before discussing results for specific systems, we remind the reader of
some basic connections between fluid structure and phase behavior.

Given an expression for the Helmholtz free energy functional for a fluid,
it is then straight forward to calculate the bulk fluid phase
behavior. The Helmholtz free energy $F$ for the bulk fluid is simply
obtained by substituting $\rho_i({\bf r})=\rho_i^{\rm b}$ into
$\mathcal{F}[\{\rho_i\}]$ and the pressure
$P$ follows as $P = -(\partial F/\partial V)_{N_1,N_2,T}$, where $V$
is the volume of the system.
Through a Legendre transformation, the Gibbs free energy $G(N_1,N_2,P,T) = 
F(N_1,N_2,V,T) + PV$ is obtained, along with the intensive 
Gibbs free energy per particle, $G/N\equiv g(x,P,T)$, where 
$x = N_2/N$ and $N=N_1+N_2$.
If, at fixed $P$ and $T$, $g(x,P,T)$ is a convex function of $x$
($g^{''}(x) > 0$) for all $x$-values, the homogeneous mixture is 
stable. If, on the other hand, there are parts with $g^{''}(x) \leq 0$,
then the system exhibits bulk fluid-fluid
phase separation and the coexisting state points are obtained by equating
the pressure and chemical potentials in the coexisting phases.\cite{andy1}
Within mean field (van der Waals-like) theories such as the present,
a fluid-fluid coexistence curve (binodal) is always accompanied by a
spinodal which is the locus in the phase diagram where the inverse
isothermal compressibility $\chi_T^{-1}=0$, -- i.e.~the locus of points at which
$(\partial^2g/\partial x^2)_{P,T}=0$. The compressibility $\chi_T$
is simply related to the $q=0$ value of the fluid static structure factor
$S(q)$,\cite{hansen:book:86} so
the spinodal can equally be defined as
the locus in the phase diagram at which $S(q=0)\rightarrow \infty$.
$S(q)$ can be expressed as a linear combination of the partial structure
factors\cite{hansen:book:86,likos:ashcroft}
in Eq.~(\ref{eq:S_q}), i.e., as a linear combination of the
functions $\hat{h}_{ij}(q)$. From
Eq.~(\ref{eq:OZ}) we see that the spinodal is given by the locus in the
phase diagram of $D(q=0)=0$. Using the RPA approximation $c_{ij}(r)=-\beta
V_{ij}(r)$ in Eq.~(\ref{Dq}), we obtain an analytic expression for
$D(q)$ and thus we have a simple way of determining if a particular
dendrimer mixture has a
spinodal and therefore exhibits bulk fluid-fluid phase
separation.\cite{andy1} 

In addition to exhibiting macro-phase separation, 
binary mixtures of Gaussian particles are capable of exhibiting
micro-phase separation.\cite{andy2} This occurs when at a wavenumber
$q=q_c>0$, there is a divergence $S(q_c)\rightarrow \infty$ (and
therefore also a divergence in $S_{ij}(q)$ at $q=q_c$). This is equivalent
to $D(q_c) \rightarrow 0$. We define the
$\lambda$-line as the locus in the phase diagram at which $D(q \neq 0)=0$.
This line has a U-shape and separates the $(\rho,x)$-plane into two regions.
Below the $\lambda$-line $D(q)>0$ for all $q$ values,\cite{andy2} whereas
above it the uniform fluid mixture is unstable.
Using the RPA approximation in Eq.~(\ref{Dq}) it is simple to locate the 
$\lambda$-line for mixtures exhibiting micro-phase separation.

\section{Bulk results}
\label{sec:results}

In this section, we display results for the bulk structure and phase
behavior of two particular binary mixtures of dendrimers, the first of
which exhibits bulk fluid-fluid phase
separation (macro-phase separation) and the
second exhibits micro-phase separation. In both these systems the
macro-phase separation (in the former) and the micro-phase separation 
(in the latter) occur at fluid
densities sufficiently high as to make simulating
these systems near to phase separation computationally expensive,
particularly since
close to the spinodal/$\lambda$-line, bulk simulations become
impractical due to the long-range nature of the fluid
correlation functions, requiring very large simulation systems to obtain
reliable results.
Thus, we compare results from the RPA DFT with simulations at lower
densities, in order to evaluate the reliability of the theory. Given that
the accuracy of the RPA approximation improves with increasing
density,\cite{habil,likos:lang,lang:jpcm:00}
the good agreement between theory and simulations that we find at lower
densities implies that the RPA is reliable for determining the phase
behavior of Gaussian mixtures at higher densities.
In our simulation results, we excluded finite size effects by performing
simulations at the same density but different total numbers of dendrimers
(20\,000 and 100\,000), that both yield identical results.

\subsection{Macro-phase separation}

\begin{figure}
  \begin{center}
  \includegraphics[width=8cm,clip]{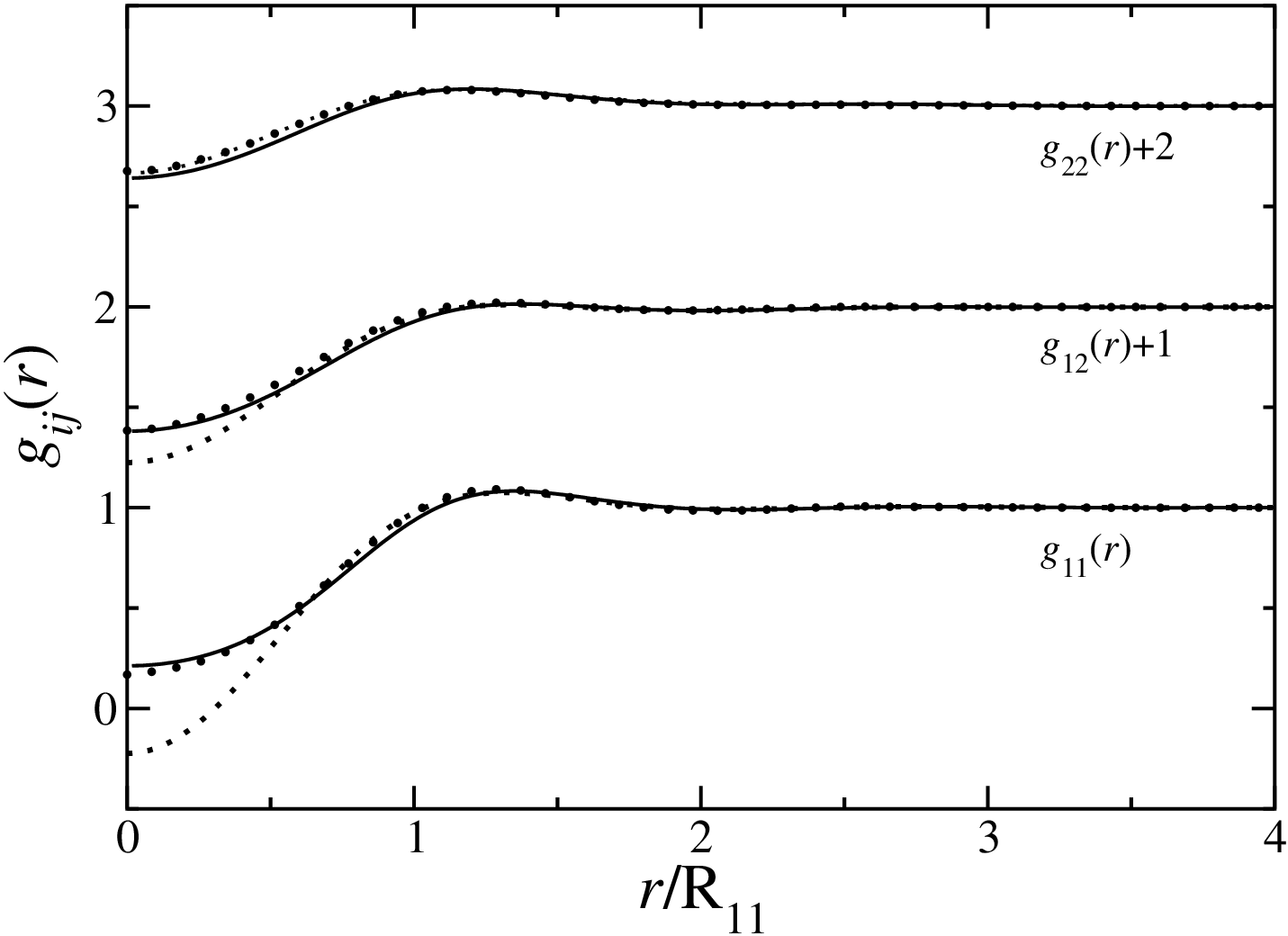}
  \end{center}
  \caption{Partial radial distribution functions $g_{ij}(r)$ for system A
  with a total density $\rho{\rm R}_{11}^3=1.49$ and concentration
  $x=0.5$. The circles are MC simulation
  results, the solid lines
  DFT test-particle results and the dotted lines results from
  the RPA closure to the OZ equations.}
  \label{macro:Gr}
\end{figure}

The first system we investigate, hereafter referred to as `system A',
consists of G4 dendrimers with $\delta=2.0\sigma$ (in the following termed
species 1), and G3 dendrimers with $\delta=3.0\sigma$ (species 2). In this
mixture the ranges of the interaction potentials ${\rm R}_{ij}$ are all
very similar, but they differ markedly in the prefactors $\epsilon_{ij}$.
The parameters of the effective pair potentials in Eq.~(\ref{eq:V_ij}) for
system A are:
\begin{eqnarray}
\epsilon_{11}&=& 11.6\, k_BT,\quad {\rm R}_{11}=\sqrt{34}\sigma\nonumber\\
\epsilon_{22}&=& 3.45\, k_BT,\quad {\rm R}_{22}=\sqrt{31}\sigma\nonumber\\
\epsilon_{12}&=& 6.4\, k_BT,\quad {\rm R}_{12}=\sqrt{33}\sigma\nonumber.
\end{eqnarray}
The pair potentials with these parameters are plotted in
Fig.~\ref{macro:pot}.
The radial distribution functions $g_{ij}(r)$ for system A are displayed
in Fig.~\ref{macro:Gr} for the total bulk density
$\rho{\rm R}_{11}^3=(\rho_1^{\rm b}+\rho_2^{\rm b}){\rm R}_{11}^3=1.49$
and concentration $x=\rho_2^{\rm b}/\rho=0.5$. We plot MC
simulation results, results obtained by inserting the RPA
closure into the OZ equations and those from using the RPA DFT via
the test-particle route. The latter route guarantees $g_{ij}(r)>0$ and is much
more reliable than the RPA OZ route. Nonetheless, the RPA OZ results still
agree well with the simulations results for $r >  {\rm R}_{11}$. It is at
small $r$ where deviations and even unphysical negative values of
$g_{ij}(r)$ appear in the RPA OZ results at lower total densities $\rho$.

\begin{figure}
  \begin{center}
  \includegraphics[width=8cm,clip]{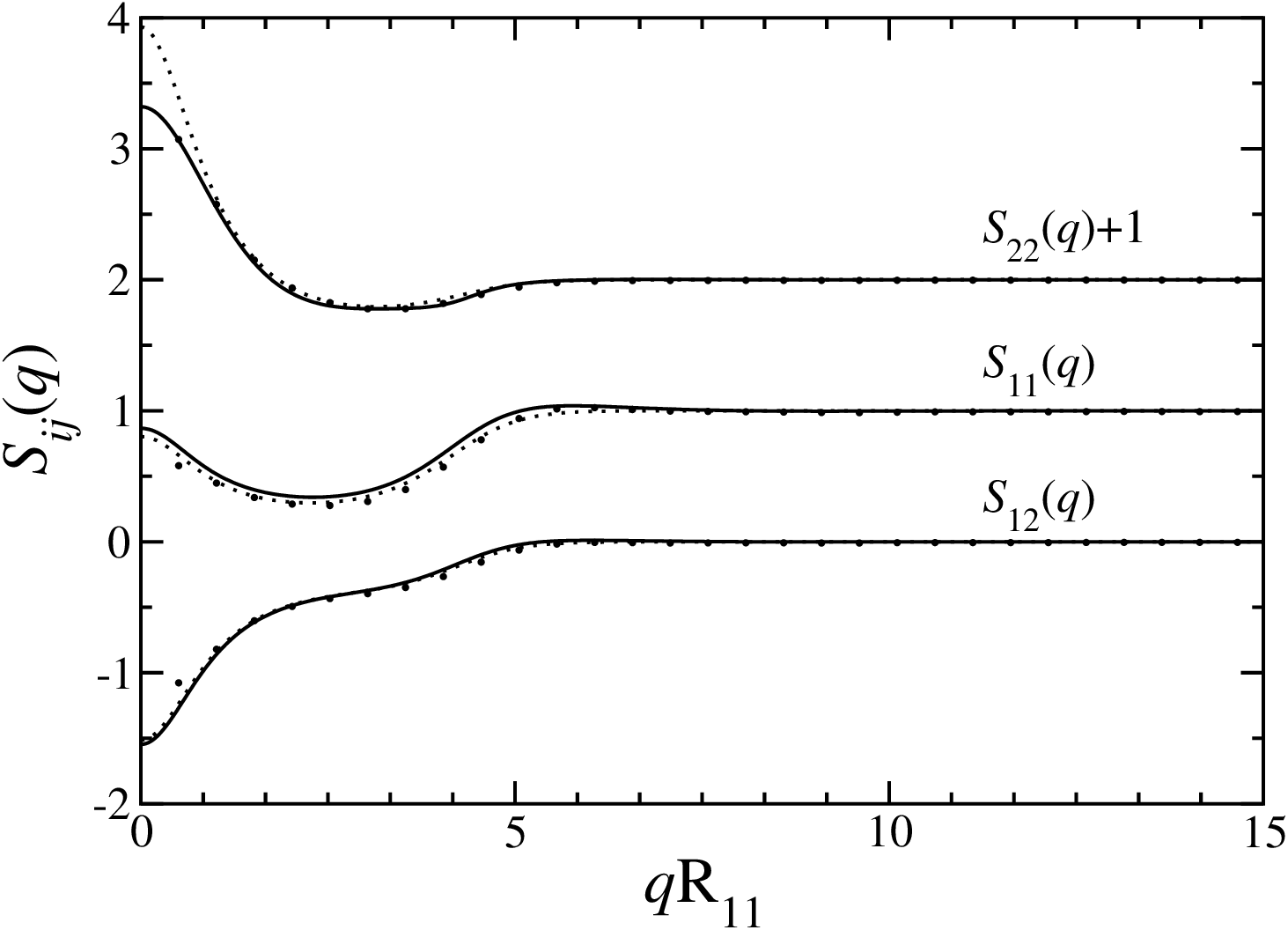}
  \end{center}
  \caption{Partial structure factors $S_{ij}(q)$ for system A
  at concentration $x=0.5$ and total density $\rho{\rm R}_{11}^3=1.49$.
  The circles are MC simulation results, the solid lines DFT test-particle
  results and the dotted lines results from using
  the RPA closure to the OZ equations.}
  \label{macro:Sq}
\end{figure}

\begin{figure}
  \begin{center}
  \includegraphics[width=8cm,clip]{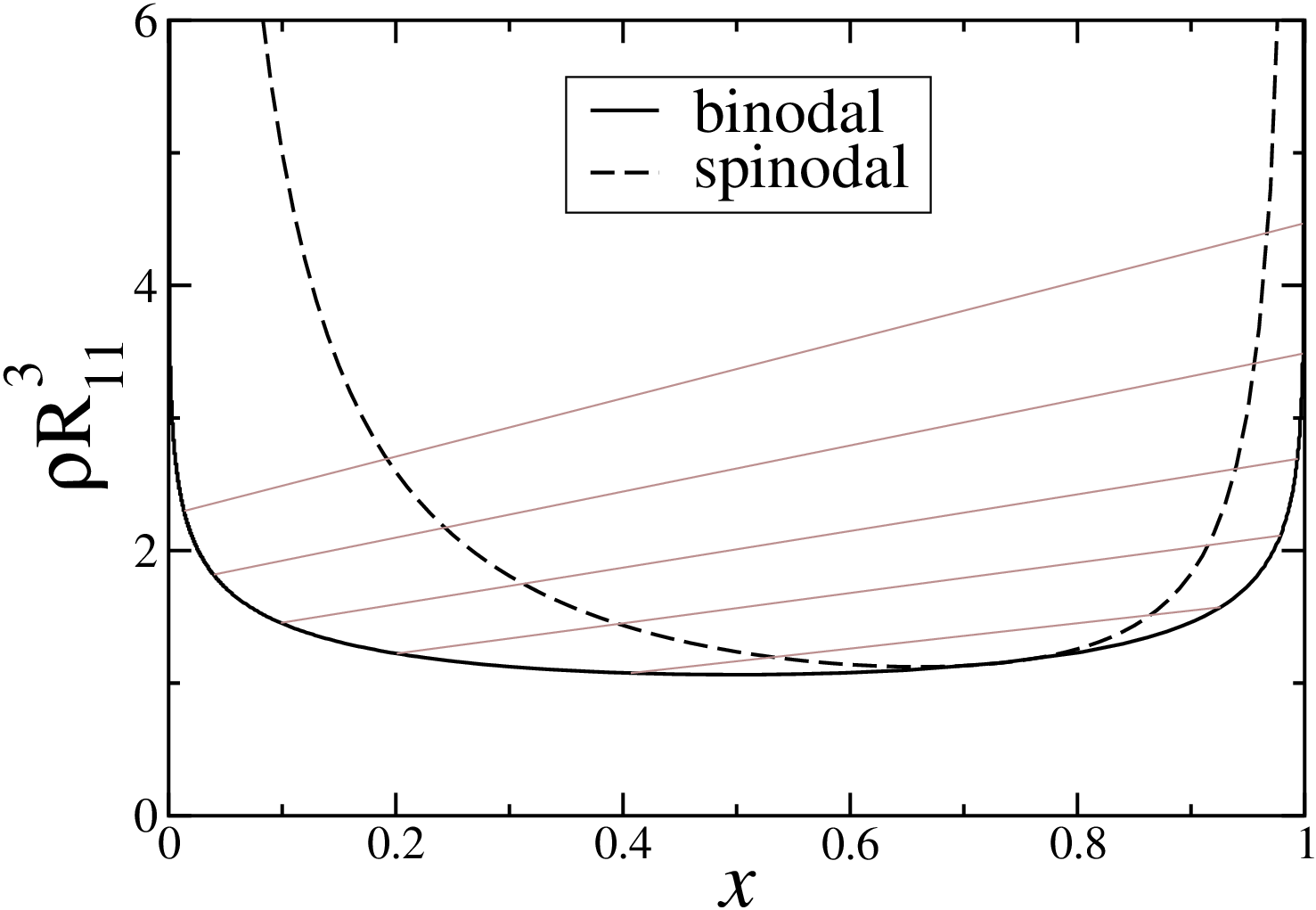}
  \end{center}
  \caption{Phase diagram for the binary dendrimer mixture exhibiting bulk
  fluid-fluid phase separation (system A), plotted in the total density, $\rho
  {\rm R}_{11}^3$, versus concentration, $x=\rho_2^{\rm b}/\rho$ plane.
  The straight lines are the tielines linking coexisting state points on the
  binodal.}
  \label{binodal}
\end{figure}

In Fig.~\ref{macro:Sq} we display the partial structure factors
$S_{ij}(q)$ for system A at the state point corresponding to the results
displayed in Fig.~\ref{macro:Gr}. We see a peak in the structure factors
$S_{11}(q)$ and $S_{22}(q)$ at $q=0$, indicating that the state point for
which these results are calculated is not too far from the spinodal.
System A exhibits bulk fluid-fluid phase separation -- i.e.~we find
a solution to the equation $D(q=0)=0$, the spinodal for this system.
In Fig.~\ref{binodal} we display both the spinodal and the binodal for
system A. These are located at rather high (monomer) densities. For
example, for $x=0.5$ the spinodal is located at a monomer density of
$\sigma^3 \rho_m=0.47$, where $\sigma$ is the monomer diameter.
At such densities the reliability of the effective pair potential for
dendrimers is
perhaps questionable. However, the influence of the
macro-phase separation is observable at lower densities, where the
structure factors display a local maximum at $q=0$ -- see
Fig.~\ref{macro:Sq}.

\subsection{Micro-phase separation}

\begin{figure}
  \begin{center}
  \includegraphics[width=8cm,clip]{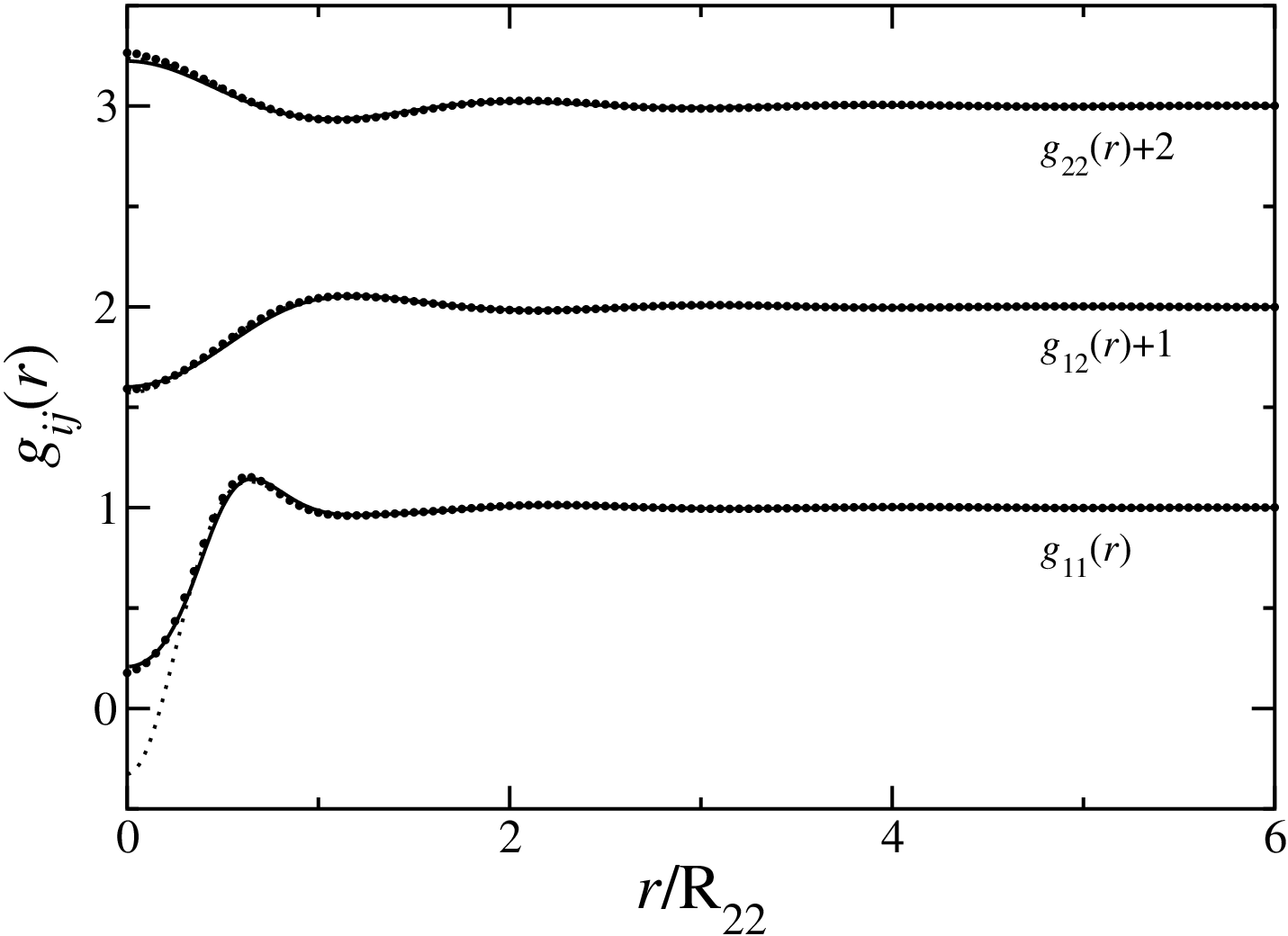}
  \end{center}
  \caption{Partial radial distribution functions $g_{ij}(r)$
  for system B
  with a total density $\rho{\rm R}_{11}^3=0.96$ and concentration
  $x=0.5$. The circles are MC simulation
  results, the solid lines
  DFT test-particle results and the dotted lines results from
  the RPA closure to the OZ equations.}
  \label{micro:Gr}
\end{figure}

The second Gaussian mixture for which we display results, `system B', is a
binary mixture of G3 dendrimers with thread length $\delta=2.0 \sigma$
(species 1) and G4 dendrimers with $\delta =4.7 \sigma$ (species 2).
Here, the 1-1 interaction is very soft and long ranged, whereas the 2-2
potential is comparatively strong and short ranged.
The parameters in the effective pair potentials [Eq.~(\ref{eq:V_ij})],
obtained from MC simulations are:
\begin{eqnarray}
\epsilon_{11}&=& 6.3\, k_BT,\quad {\rm R}_{11}=\sqrt{21}\sigma\nonumber\\
\epsilon_{22}&=& 3.1\, k_BT,\quad {\rm R}_{22}=\sqrt{100}\sigma\nonumber\\
\epsilon_{12}&=& 3.2\, k_BT,\quad {\rm R}_{12}=\sqrt{52}\sigma\nonumber.
\end{eqnarray}
The pair potentials with these parameters are plotted in
Fig.~\ref{micro:pot}.
The radial distribution functions $g_{ij}(r)$ for system B are displayed
in Fig.~\ref{micro:Gr} for the total bulk density
$\rho{\rm R}_{11}^3=0.96$ and concentration $x=0.5$. We plot MC
simulation results, results obtained by inserting the RPA
closure into the OZ equations and those from using the RPA DFT via
the test-particle route. We find good agreement between the MC and DFT
results. The functions $g_{ij}(r)$ decay rather slowly in an
oscillatory manner, with wavelength $\lambda \simeq 2{\rm R}_{22}$.
The presence of slowly decaying oscillations 
indicates that this state point is not too far
in the phase diagram from the $\lambda$-line.
In fact, it turns out that the density
$\rho{\rm R}_{11}^3=0.96$, corresponds, at $x = 0.5$, to 
$0.76\rho_{\lambda}(x=0.5)$, where $\rho_{\lambda}(x)$ is the 
$x$-dependent value of the total density on the $\lambda$-line.
Moreover, the fact that $g_{22}(0)>1$ indicates that the dendrimers of
species 2 prefer complete overlap, a further indication of the propensity
towards microphase ordering; recall the potential $V_{22}(r)$ is purely
repulsive.

\begin{figure}
  \begin{center}
  \includegraphics[width=8cm,clip]{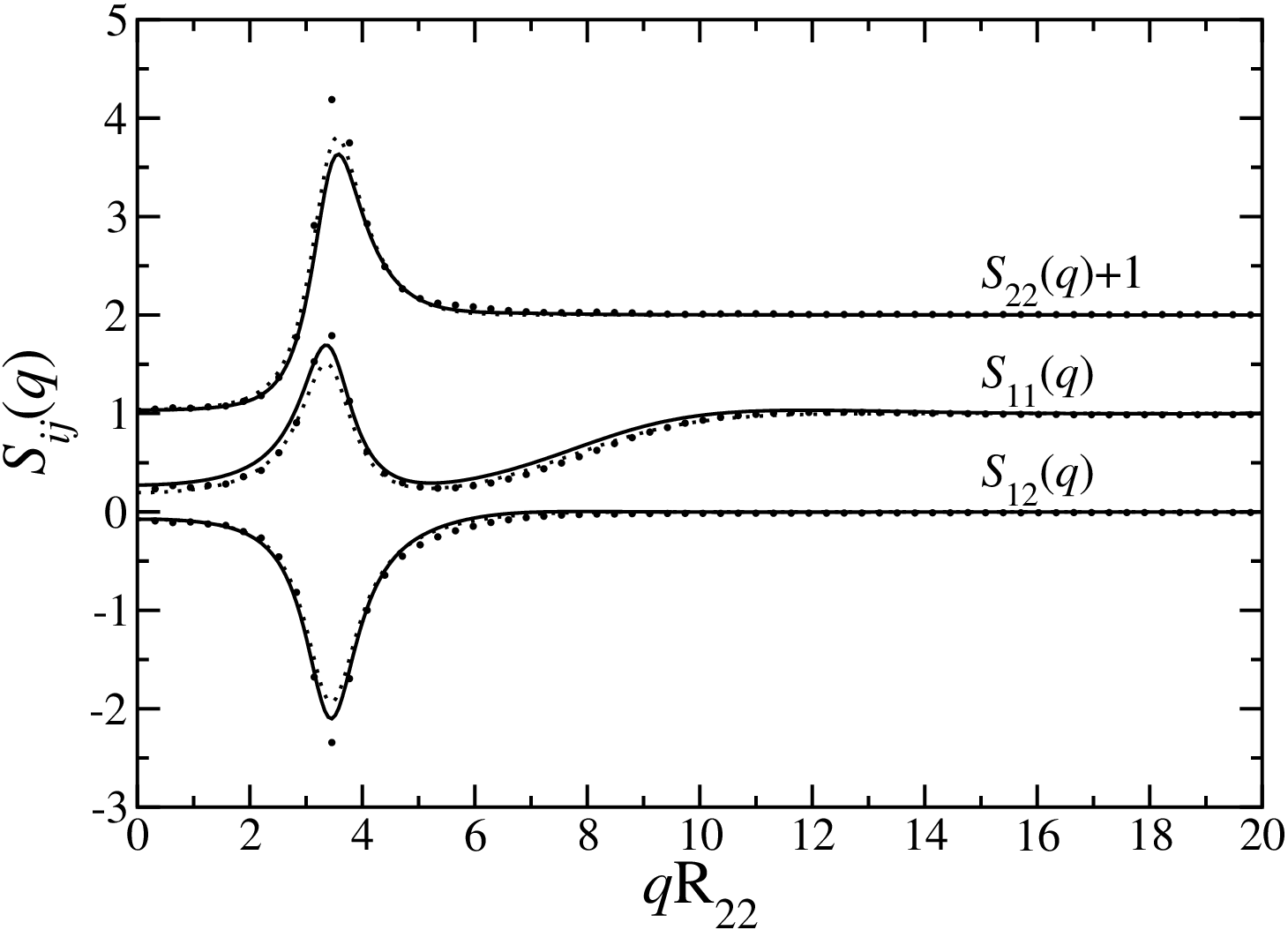}
  \end{center}
  \caption{Partial structure factors $S_{ij}(q)$ for system B
  at concentration $x=0.5$ and total density $\rho{\rm R}_{11}^3=0.96$.
  The circles are MC simulation results, the solid lines DFT test-particle
  results and the dotted lines results from using
  the RPA closure to the OZ equations.}
  \label{micro:Sq}
\end{figure}

\begin{figure}
  \begin{center}
  \begin{minipage}[t]{8cm}
  \includegraphics[width=8cm,clip]{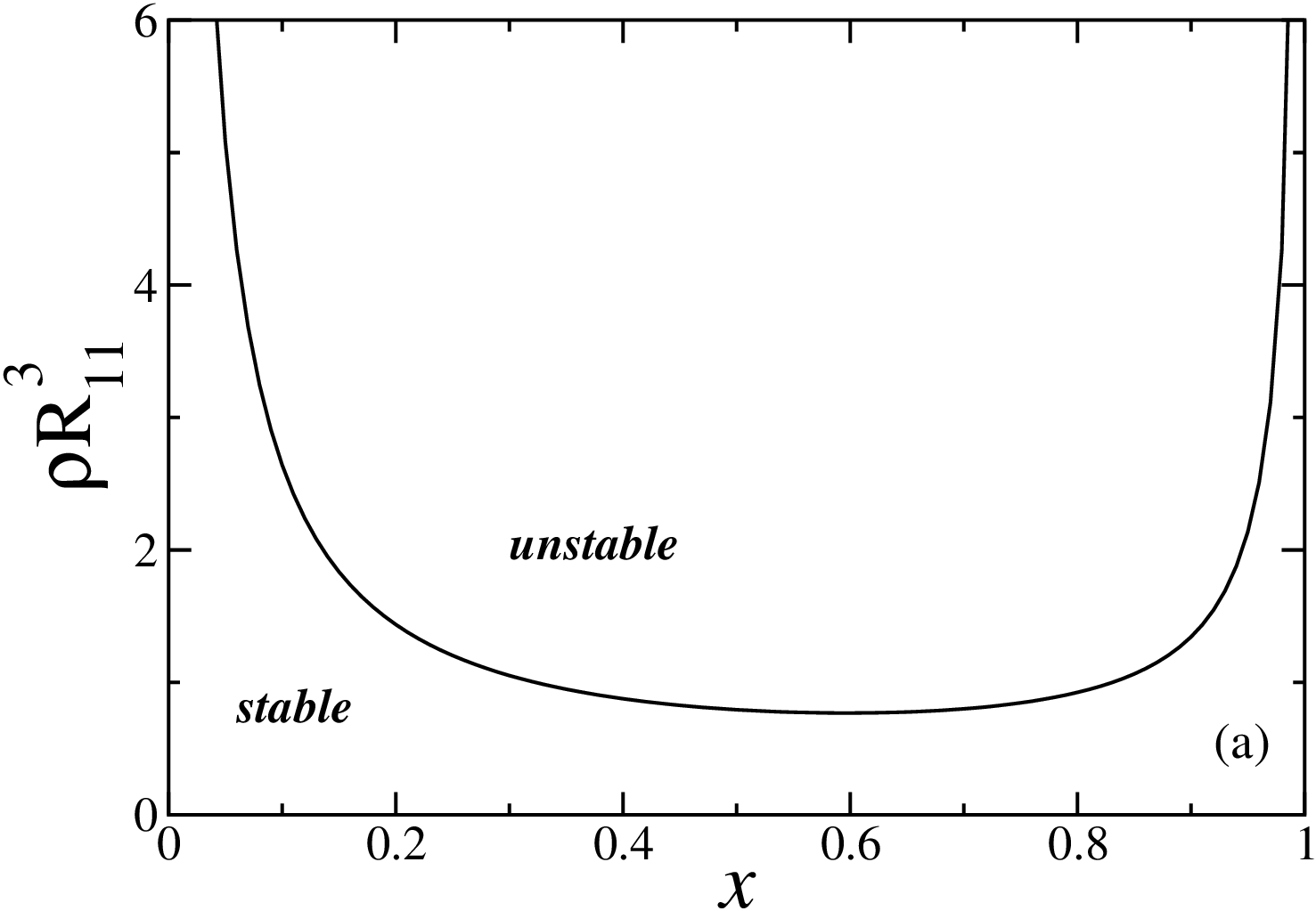}
 \end{minipage}
\hfill
   \begin{minipage}[t]{8cm}
  \includegraphics[width=8cm,clip]{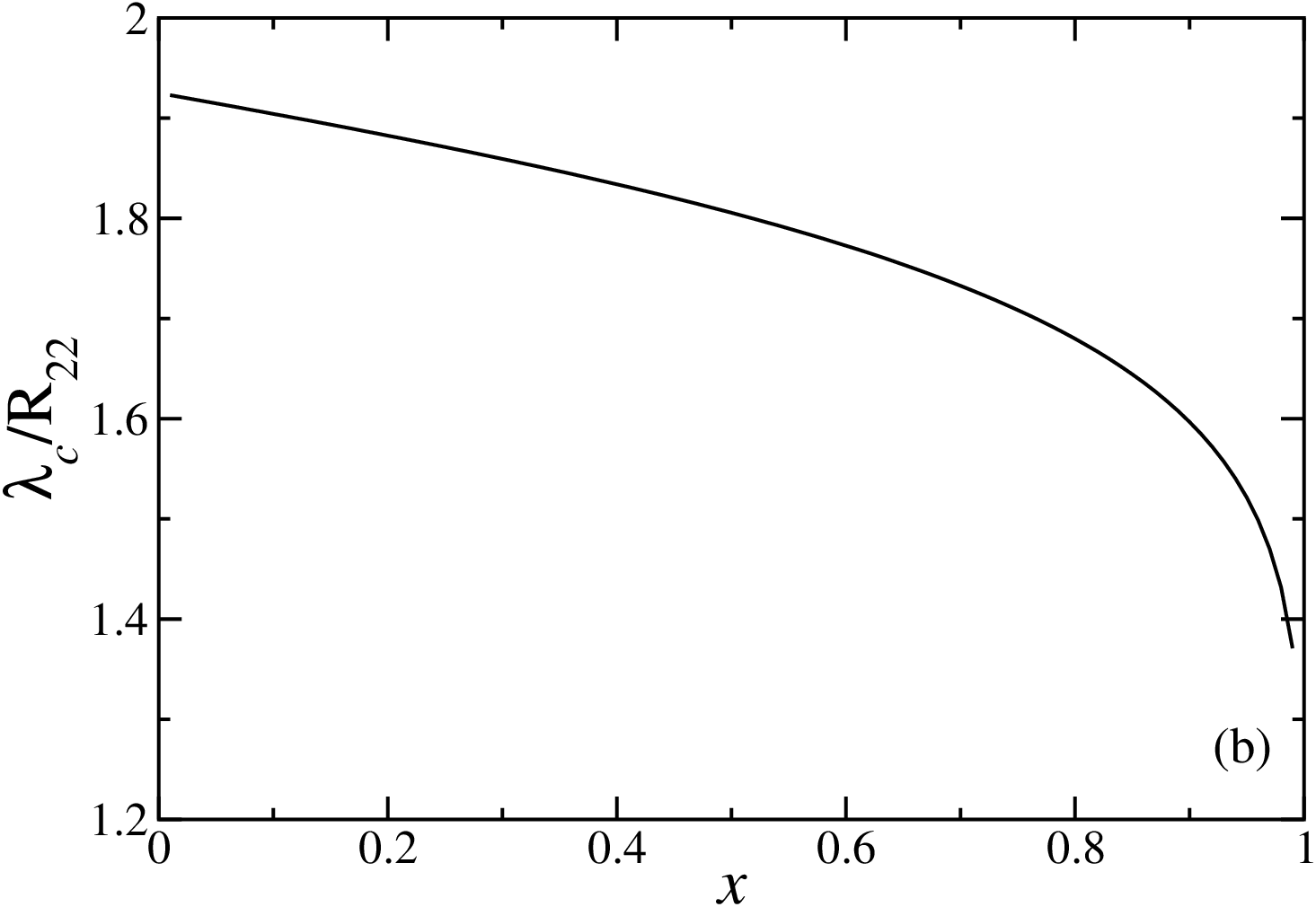}
\end{minipage}
   \end{center}
  \caption{(a) Stability `phase diagram' for the binary dendrimer mixture
  exhibiting micro-phase separation, plotted in the total density, $\rho
  {\rm R}_{11}^3$, versus concentration, $x=\rho_2^{\rm b}/\rho$ plane.
  The solid line is the $\lambda$-line. (b) The wavelength 
  $\lambda_c = (2\pi)/q_c$ of the instability on the $\lambda$-line,
  where $q_c$ is the wavenumber for which $D(q_c) = 0$.}
  \label{lambdaline}
\end{figure}

In Fig.~\ref{micro:Sq} we display the partial structure factors
$S_{ij}(q)$ for system B at the state point corresponding to the results
displayed in Fig.~\ref{micro:Gr}. There is good agreement between the MC
simulation results and the DFT results.
We see a peak in the structure factors
$S_{11}(q)$ and $S_{22}(q)$ at $q=q_c \simeq 3.5{\rm R}_{22}^{-1}$,
indicating the propensity towards micro-phase ordering.
System B exhibits a $\lambda$-line, i.e., we find
a solution to the equation $D(q=q_c>0)=0$.
In Fig.~\ref{lambdaline}(a) we display the $\lambda$-line for
system B and
in Fig.~\ref{lambdaline}(b) the wavelength $\lambda_c$ of the instability,
$\lambda_c = 2\pi/q_c$, for the points along the $\lambda$-line.
The $\lambda$-line 
is located at rather high densities: its minimum density $\rho{\rm
R}_{11}^3=1.23$ corresponds to a monomer density of $\sigma^3
\rho_m=0.63$. The length scale of the instability,
$\lambda_c$, is somewhat smaller than 
$2{\rm R}_{22}$ and is rather insensitive to the precise
location on the instability line. This indicates that the length scale of the 
$\lambda$-line instability is mainly determined
by the larger particles, whose size
and interaction range is indeed ${\rm R}_{22}$. However, it must
be emphasized that the emergence of this inherent instability is
a characteristic of the {\it whole mixture} and it requires the
presence of both species with the associated interactions between
like and unlike species.

\section{Confined systems}
\label{sec:confined}

We determine the inhomogeneous fluid density profiles of systems A and
B, from the previous section, when these binary dendrimer mixtures are
confined in a spherical cavity. 
Much is known about the behavior of Gaussian mixtures at planar
walls\cite{andy1b} and in spherical
cavities.\cite{archerJPCM2005,archerJPCM2005_LMC} However, this
previous work relied (essentially) on a guess for the form of the external
potentials. Here, we determine effective external potentials between the
dendrimers and the wall explicitly.
The potential between the individual
dendrimer monomers and the cavity
wall is assumed to be a hard interaction (i.e.~the potential of the
monomer is zero inside the cavity and infinite outside the cavity). We
calculate an effective potential between the centre of mass of the
dendrimers and the wall in a manner similar to that used to derive the
dendrimer-dendrimer effective pair potentials -- i.e.~we perform
biased MC simulations of a single monomer-resolved dendrimer near a planar
wall where the potential of the monomer is zero for $z>0$ ($z$ is the
Cartesian axis perpendicular to the wall) and infinite for $z<0$.
This planar effective wall potential has the following Yukawa form:
\begin{equation}
V_i^{\rm ext}(z)=\left\{\begin{array}{l@{\qquad}l}
\frac{\epsilon_{i} {\rm R}_{11} \exp[-z/{\rm R}_{i}]}{z} & 
\mbox{for}\quad z > 0 \\ \infty & \mbox{otherwise}.
\end{array} \right. 
\label{eq:planar_wall_pot}
\end{equation}
For mixture A, the parameters are $\epsilon_1=150 k_BT$, ${\rm
R}_1=0.34{\rm R}_{11}$, $\epsilon_2=70 k_BT$ and
${\rm R}_2=0.39{\rm R}_{11}$.
For mixture B, the parameters are $\epsilon_1=68 k_BT$, ${\rm
R}_1=0.39{\rm R}_{11}$, $\epsilon_2=150 k_BT$ and
${\rm R}_2=0.89{\rm R}_{11}$.
From Eq.~(\ref{eq:planar_wall_pot}) we may then construct the external
potential corresponding to a spherical cavity of radius ${\rm R}_{\rm
wall}$:
\begin{equation}
V_i^{\rm ext}(r)=\left\{\begin{array}{l@{\qquad}l}
\frac{\epsilon_{i} {\rm R}_{11}
\exp[-({\rm R}_{\rm wall}-r)/{\rm R}_{i}]}{({\rm R}_{\rm wall}-r)} & 
\mbox{for}\quad r < {\rm R}_{\rm wall} \\ \infty & \mbox{otherwise}.
\end{array} \right. 
\end{equation}
Of course, this will only provide a reliable approximation to the cavity
potential when ${\rm R}_{\rm wall} \gg {\rm R}_{ij}$.

\begin{figure}
  \begin{center}
  \includegraphics[width=8cm,clip]{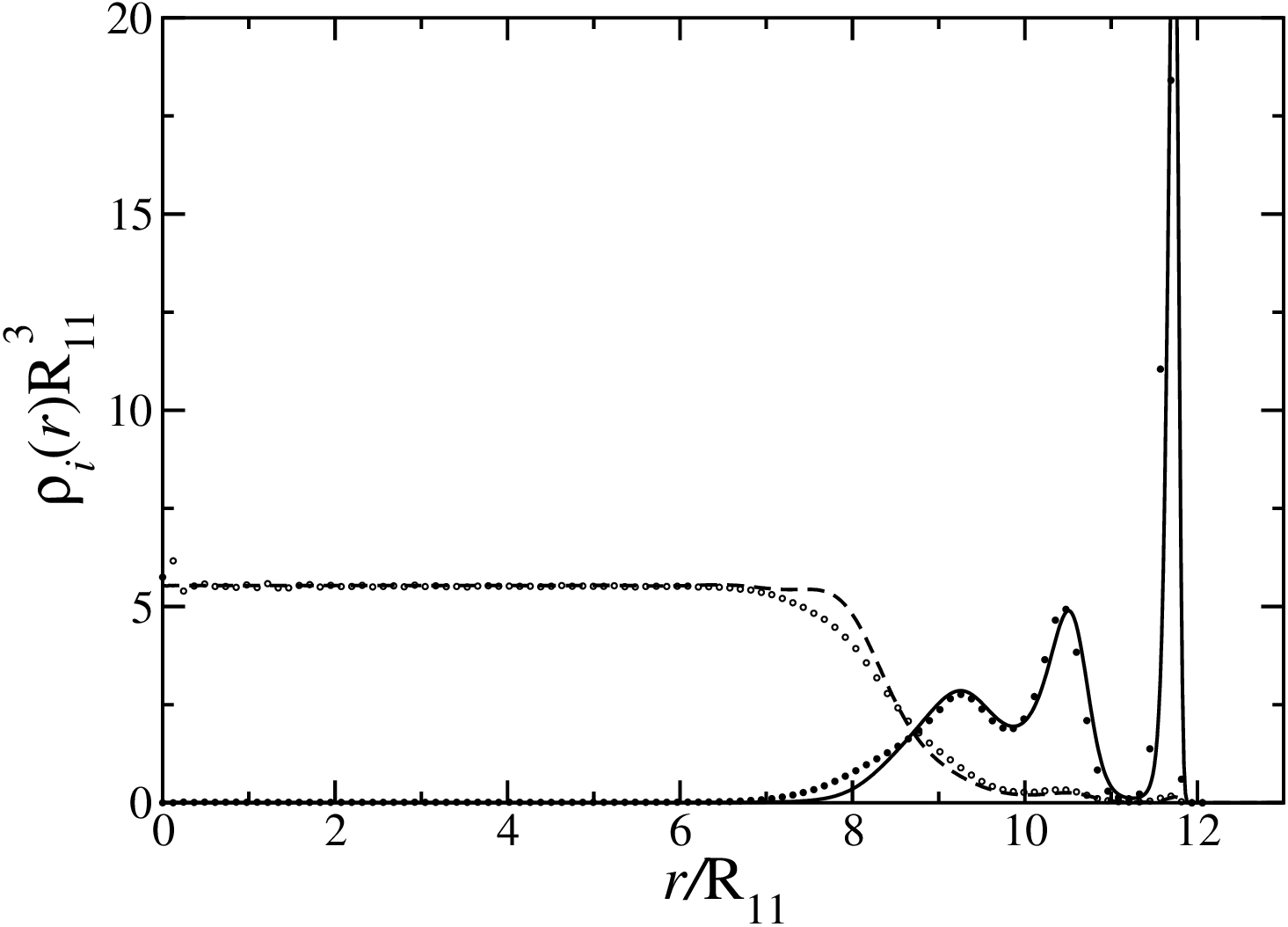}
  \end{center}
  \caption{Density profiles for system A with $N_1=15\,000$ particles of
  species 1 and $N_2=15\,000$ particles of species 2 confined in a cavity
  of radius ${\rm R}_{\rm wall}/{\rm R}_{11}=12.2$. The solid line is the
  DFT result for species 1 and the dashed line species 2. The symbols are
  MC simulation results.}
  \label{conf:macro}
\end{figure}

In Fig.~\ref{conf:macro} we display the fluid density profiles for
system A with $N_1=15000$ particles of species 1 and $N_2=15000$ particles
of species 2 confined in a cavity with
${\rm R}_{\rm wall}/{\rm R}_{11}=12.2$. The average total density in the
cavity corresponds to a state point well inside the fluid-fluid demixing
binodal. This phase separation is manifest in the fluid one body density
profiles. There is preferential adsorption of species 1 at the walls
of the cavity and there are almost exclusively just particles of species 2
for $r \lesssim 8\,{\rm R}_{11}$,
in the centre of the cavity. Which of the
two species is preferentially adsorbed by the cavity
walls is a rather fine balance, determined by an interplay of the various
external wall potential parameters. Slightly changing one of these
parameters can lead to adsorption of the other species at the wall.

\begin{figure}
  \begin{center}
  \includegraphics[width=8cm,clip]{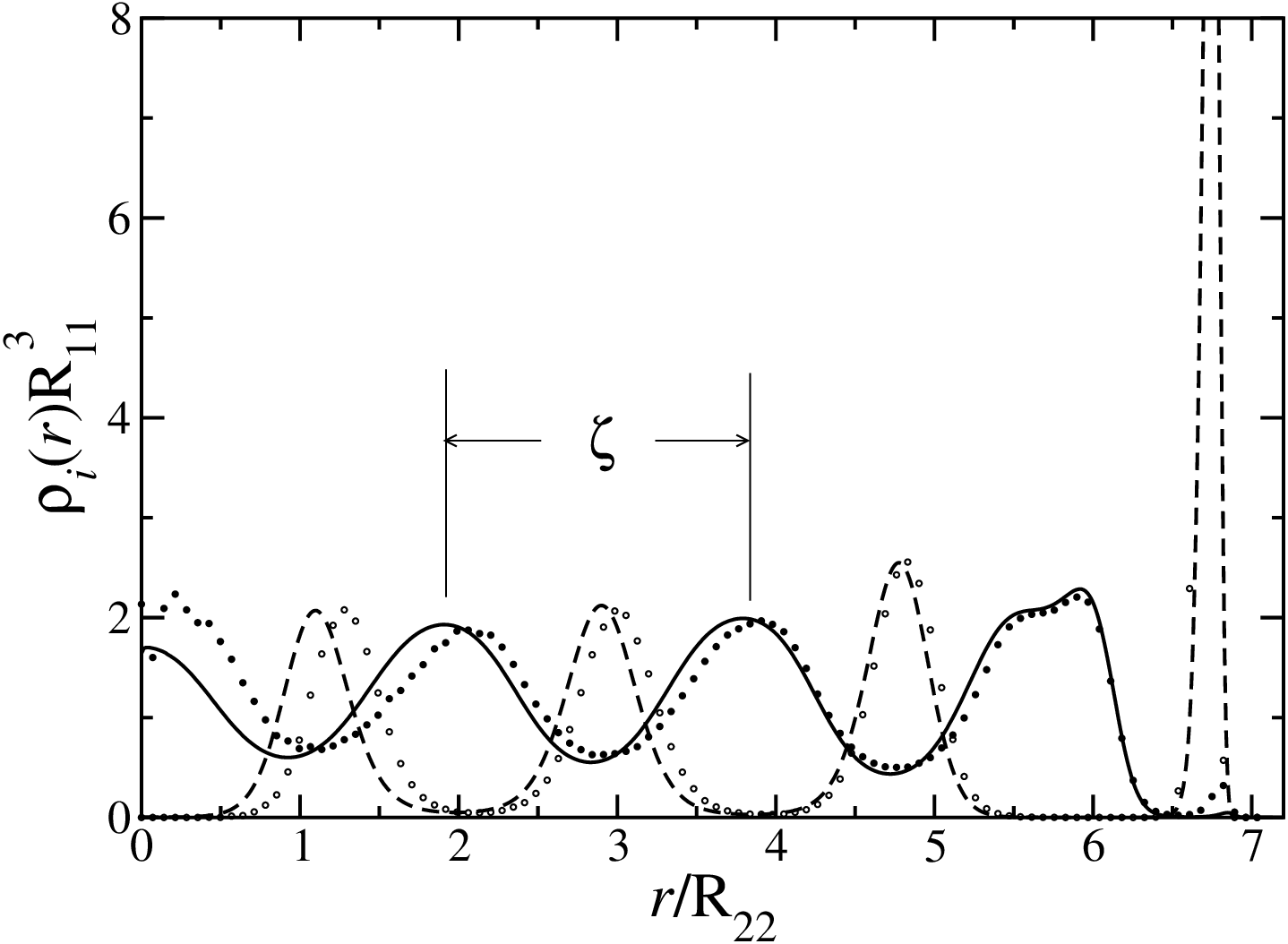}
  \end{center}
  \caption{Density profiles for system B with $N_1=15\,000$ particles of
  species 1 and $N_2=15\,000$ particles of species 2 confined in a cavity
  of radius ${\rm R}_{\rm wall}/{\rm R}_{22}=7.2$. The solid line is the
  DFT result for species 1 and the dashed line species 2. The symbols are
  MC simulation results.}
  \label{conf:micro}
\end{figure}

In Fig.~\ref{conf:micro} we display the fluid density profiles for
system B with $N_1=15000$ particles of species 1 and $N_2=15000$ particles
of species 2 confined in a cavity with
${\rm R}_{\rm wall}/{\rm R}_{22}=7.2$. The average total density in the
cavity corresponds to a state point inside the $\lambda$-line.
As a consequence, we see rather striking `onion-like' ordering in the
cavity -- alternating layers rich in the two different species of
particles. 
The DFT and MC simulation results agree well; the DFT is able to capture
the details of the highly structured profiles, although a slight phase
shift in the oscillations in the profiles is noticeable.
This ordering has its origins in the existence of the
inherent $\lambda$-line instability of the bulk system. Indeed,
as can be seen in Fig.~\ref{conf:micro}, the length scale $\zeta$
that characterizes the onion structure is roughly 
$\zeta \cong 2{\rm R}_{22}$,
precisely as the critical wavelength $\lambda_c$ shown in
Fig.~\ref{lambdaline}(b).
Under planar confinement, mixture B would have formed alternating layers
of species 1-rich and species 1-poor lamellae instead.\cite{andy2}

\section{Conclusions}
\label{sec:conclusions}

We have investigated mixtures of dendrimers, finding either bulk
fluid-fluid demixing or micro-phase separation, depending on the choice of
the dendrimers' generation numbers and architecture.
Here, in contrast to previous studies concerning micro-phase separation,
we do {\it not} employ arbitrary Gaussian potentials but those obtained
from monomer-resolved simulations of dendrimers.
Furthermore, we observe pattern formation under confinement, finding a
strong sensitivity with respect to the form of the wall potential.
This work demonstrates, therefore, that the rich phenomenology
encountered in Gaussian mixtures can be realized by employing
dendrimers as soft colloids with tunable repulsions. It would be
desirable to synthesize suitable mixtures of dendrimers, along the
lines put forward in thus work, with an inherent $\lambda$-type
instability. Under planar confinement, these would organize in
lamellae, which opens the possibility of using dendrimers as
lubrication agents, for instance. Another very interesting question
is the nature of the phase or phases that form inside the unstable
region enclosed by the $\lambda$-line. A rich variety of one-, two-, or
three-dimensionally modulated structures (cylinders, lamellae, crystals),
similar to that encountered in block copolymer blends or in
ternary oil/water/surfactant mixtures could be expected here as well.
We plan to return to this problem in the future.

\acknowledgments

This work was funded in part by the Deutsche Forschungsgemeinschaft
(DFG). A.J.A. acknowledges the support of EPSRC under Grant
No.~GR/S28631/01.


\begin{thebibliography}{99}

\bibitem{newcome:96} G.~R.~Newcome, C.~N.~Moorefield, and
F.~V{\"o}gtle, {\it Dendritic Molecules: Concepts, Synthesis,
Perspectives} (Wiley-VCH: Weinheim 1996).

\bibitem{fischer:99} M.~Fischer and F.~V{\"o}gtle, 
Angew. Chem. Int. Ed. {\bf 38}, 884 (1999).

\bibitem{bosman:99} A.~W.~Bosman, H.~M.~Janssen, and E.~W.~Meijer,
Chem. Rev. {\bf 99}, 1665 (1999).

\bibitem{tomalia:01} {\it Dendrimers and Other Dendritic Polymers},
ed.\ by J.~M.~J.~Fr{\'e}chet and D.~A.~Tomalia (Wiley-VCH: Weinheim 2001).

\bibitem{likos:04} M.~Ballauff and C.~N.~Likos, 
Angew. Chem. Int. Ed. {\bf 43}, 2998 (2004).

\bibitem{lee} I.~Lee, B.~T.~Athey, A.~W.~Wetzel, W.~Meixner, and
J.~R.~Baker, Macromolecules {\bf 35}, 4510 (2002).

\bibitem{boas:04}
U.~Boas and P.~M.~H.~Heegaard,
Chem. Soc. Rev. {\bf 33}, 43 (2004).

\bibitem{quintana:02}
A.~Quintana, E.~Raczka, L.~Piehler, I.~Lee, A.~Myc, I.~Majoros, 
A.~K.~Patri, T.~Thomas, J.~Mule, and J.~R.~Baker,
Pharm. Res. {\bf 19}, 1310 (2002).

\bibitem{kohle:03}
P.~Kolhe, E.~Misra, R.M.~Kannan, S.~Kannan, and M.~Lieh-Lai,
Int. J. Pharm. {\bf 259}, 143 (2003).

\bibitem{choi:05}
Y.~Choi and J.~R.~Baker, Cell Cycle {\bf 4}, 669-671 (2005).

\bibitem{light-harvesting}
A.~Adronov and J.~M.~J.~Frechet,
Chem. Commun. {\bf 18}, 1701 (2000). 

\bibitem{goessl:02} I.~G{\"o}ssl, L.~Shu, A.~D.~Schl{\"u}ter, 
and J.~P.~Rabe, J. Am. Chem. Soc. {\bf 124}, 6860 (2002).

\bibitem{habil}
C.~N.~Likos, Phys. Rep. {\bf 348}, 267 (2001).

\bibitem{ingo:jcp:04}
I.~O.~G{\"o}tze, H.~M.~Harreis and C.~N.~Likos, J. Chem. Phys., {\bf 120},
7761 (2004).

\bibitem{likos:rosenfeldt:02}
C.~N.~Likos, S.~Rosenfeldt, N.~Dingenouts, M.~Ballauff, P.~Lindner,
N.~Werner, and F.~V{\"o}gtle, J. Chem. Phys. {\bf 117}, 1869 (2002).

\bibitem{ingo:macrom}
I~.O.~G{\"o}tze and C.~N.~Likos, Macromolecules {\bf 36}, 8189 (2003).

\bibitem{GotzeLikosJPCM2005}
I.~O.~G{\"o}tze and C.~N.~Likos, J. Phys.: Condens. Matter {\bf 17}, S1777
(2005).

\bibitem{lang:jpcm:00}
A.~Lang, C.~N.~Likos, M.~Watzlawek and H.~L{\"o}wen,
J. Phys.: Condens. Matter {\bf 12}, 5087 (2000).

\bibitem{ard:pre:00}
A.~A.~Louis, P.~G.~Bolhuis and J.-P.~Hansen, Phys. Rev. E {\bf 62}, 7961
(2000).

\bibitem{andy1}
A.~J.~Archer and R.~Evans, Phys. Rev. E {\bf 64}, 041501 (2001).

\bibitem{andy2}
A.~J.~Archer, C.~N.~Likos and R.~Evans, J. Phys.: Condens. Matter {\bf 16},
L297 (2004).

\bibitem{sear:99}
R.~P.~Sear and W.~M.~Gelbart, J. Chem. Phys. {\bf 110}, 4582 (1999).

\bibitem{evans}
R.~Evans, Adv. Phys. {\bf 28}, 143 (1979);
R.~Evans, in {\it Fundamentals of Inhomogeneous Fluids}, ed.\ by
D.~Henderson (Dekker, New York, 1992).

\bibitem{hansen:book:86}
J.-P.~Hansen and I.~R.~MacDonald, {\it Theory of Simple Liquids},
second edition (Academic: New York, 1986).

\bibitem{andy1b}
A.~J.~Archer and R.~Evans, J. Phys.: Condens. Matter {\bf 14}, 1131 (2002).

\bibitem{likos:lang}
C.~N.~Likos, A.~Lang, M.~Watzlawek and H.~L\"owen, Phys. Rev. E {\bf 63},
031206 (2001). 

\bibitem{acedo} L.~Acedo and A.~Santos, Phys. Lett. A {\bf 323}, 427 (2004).

\bibitem{archerJPCM2005}
A.~J.~Archer, J. Phys.: Condens. Matter {\bf 17}, 1405 (2005).

\bibitem{archerJPCM2005_LMC}
A.~J.~Archer, J. Phys.: Condens. Matter {\bf 17}, S3253 (2005).

\bibitem{likos:ashcroft}
C.~N.~Likos and N.~W.~Ashcroft, J. Chem. Phys. {\bf 97}, 9303 (1992).

\end{thebibliography}
\end{document}